\documentclass[11pt]{article}
\usepackage{blois,epsfig}
\usepackage[latin1]{inputenc}
\usepackage[OT1]{fontenc}
\usepackage{bm}
\def\Journal#1#2#3#4{{#1} {\bf #2}, #3 (#4)}

\def\PLB{{\em Phys. Lett.}  B}
\def\PRL{\em Phys. Rev. Lett.}

\def\Pramana {\em Pramana - J. Phys.}
\def\Plasmas {\em Phys. of Plasmas}
\def\JHEP{\em JHEP}
\def\GravCosm{\em Gravitation and Cosmology}
\def\JPAMT{\em J.Phys.A: Math. Theor.}
\def\JPAMG{\em J.Phys.A: Math. Gen.}
\def\AFdB{\em Ann. Fond. Louis Broglie}
\def\PRC{{\em Phys. Rev.} C}
\def\RusPhysJ{\em Russ. Phys. J.}
\def\PAN{\em Phys. Atom. Nucl.}

\def\be{\begin{equation}}
\def\ee{\end{equation}}
\def\bea{\begin{eqnarray}}
\def\eea{\end{eqnarray}}

\begin{document}
\vspace*{4cm}
\title{NEUTRINOS, ELECTRONS AND MUONS IN  ELECTROMAGNETIC FIELDS AND MATTER:
THE METHOD OF EXACT SOLUTIONS}

\author{K.A. Kouzakov$^\dag$ and A.I. Studenikin$^\ddag$}
\address{$^\dag$Department of Nuclear Physics and Quantum Theory of Collisions and $^\ddag$Department of
Theoretical Physics, Moscow State University, 119992 Moscow, Russia}

%

\maketitle\abstracts{We present a powerful method for exploring various processes in the presence of strong
external fields and matter. The method implies utilization of the exact solutions of the modified Dirac
equations which contain the effective potentials accounting for the influences of external electromagnetic
fields and matter on particles. We briefly discuss the basics of the method and its applications to studies of
different processes, including a recently proposed new mechanism of radiation by neutrinos and electrons moving
in matter (the spin light of the neutrino and electron). In view of a recent ``prediction'' of an
order-of-magnitude change of the muon lifetime under the influence of an electromagnetic field of a CO$_2$
laser, we revisit the issue and show that such claims are nonrealistic.}

\section{Introduction}

The problem of particles' interactions under the influence of external electromagnetic fields and matter is one
of the important topics in particle physics. Besides the possibility for better visualization of fundamental
properties of particles and their interactions when they are influenced by external conditions, the interest to
this problem is also stimulated by astrophysical and cosmological applications, where strong electromagnetic
fields and dense matter may play important roles. There are well established methods for such kind of
investigations that have a long-standing history.

In particular, the method of exact solutions of quantum equations, which is based on a Furry representation
\cite{FurPR51} of QED, is widely used in studies of particles' interactions in external electromagnetic fields.
In this technique, the evolution operator $U_{F}(t_1, t_2)$, which determines the matrix element of the process,
is presented in the usual form
\begin{equation}
U_{F} (t_1, t_2)= T exp \left[-i \int \limits_{t_1}^{t_2}j_{\mu}(x) A^{\mu}{d}x \right],
\end{equation}
where $A_{\mu}(x)$ is the quantized part of the potential corresponding to the radiation field, which is
accounted for within the perturbation-series techniques. At the same time, the electron (a charged particle)
current is presented as
\begin{equation}
j_{\mu}(x)={e \over 2}\left[\overline \Psi_e \gamma _{\mu}, \Psi_e \right],
\end{equation}
where $\Psi_e$ are the exact solutions of the Dirac equation for an electron in the presence of an external
electromagnetic field given by the classical non-quantized potential $A_{\mu}^{ext}(x)$:
\begin{equation}\label{D_eq_QED}
\left\{ \gamma^{\mu}\left(i\partial_{\mu} -eA_{\mu}^{cl}(x)\right) - m_e \right\}\Psi_e (x)=0.
\end{equation}
Note that within this approach the interaction of charged particles with an external electromagnetic field is
taken into account exactly while the radiation field is allowed for by perturbation-series expansion techniques.
That is why the method discussed is often referred to as the ``method of exact solutions''. The detailed
discussion on foundations of this method and its applications to different processes such as, for instance, the
synchrotron radiation by an electron in magnetic fields can be found
in~\cite{SokTerSynRad68,ritus87,nikishov87}. Many results, which ate very important for astrophysical
applications, have been obtained within the discussed method when considering the neutron beta-decay in a
constant magnetic field. These studies have been started in~\cite{Kor64TerLysKor65}. Short reviews on the
studies of beta-decay and the related cross symmetric processes in strong magnetic fields can be found
in~\cite{ShiStuP05KouStuPRC05}.

\section{Neutrino and electron quantum states in matter}

Recently we have applied the ``method of exact solutions'' for treating different interactions of neutrinos and
electrons in the presence of matter
\cite{StuTerPLB05GriStuTerPLB05GriStuTerG&C05,StuJPA06StuAFdB06,GriShiStuTerTroG&C08_GriStuTerTroShiFizJ07} (see
also \cite{StuJPA08}). The developed method is based on the use of the exact solutions of the modified Dirac
equations that include effective matter potentials. It has been demonstrated how this method works in
application to different quantum processes that can take part in the presence of matter. In particular, we have
elaborated the quantum theory of the new type of electromagnetic radiation that can be emitted by a
neutrino~\cite{StuTerPLB05GriStuTerPLB05GriStuTerG&C05} and an
electron~\cite{StuJPA06StuAFdB06,GriShiStuTerTroG&C08_GriStuTerTroShiFizJ07} while these particles move in dense
matter (spin light of a neutrino, $SL\nu$, and an electron, $SL\nu$). Note that $SL\nu$ in matter was first
considered in \cite{LobStuPLB03} on the basis of a quasiclassical treatment.

As it was shown in \cite{StuTerPLB05GriStuTerPLB05GriStuTerG&C05} $^{,}$
\cite{StuJPA06StuAFdB06,GriShiStuTerTroG&C08_GriStuTerTroShiFizJ07}  (see also \cite{StuJPA08}) in the case of
the standard model interactions of the electron neutrinos and electrons with matter composed of neutrons, the
corresponding modified Dirac equations for each of the particles can be written in the following form:
\begin{equation}\label{new_e}
\left\{ i\gamma_{\mu}\partial^{\mu}-\frac{1}{2} \gamma_{\mu}(c_l+\gamma_{5}){\widetilde f}^{\mu}-m_l
\right\}\Psi^{(l)}(x)=0,
\end{equation}
where for a neutrino $m_l=m_\nu$ and $c_l=c_{\nu}=1$, whereas for an electron $m_l=m_e$ and
$c_l=c_e=1-4\sin^{2}\theta_{W}$. For unpolarized matter $\widetilde{f}^{\mu}=\frac{G_{F}}{\sqrt{2}}(n_n,n_n{\bf
v}),$ $n_n$ and $\mathbf v$ are, respectively, the neutron number density and average speed. The solutions of
these equations are
\begin{equation}
\Psi^{(l)}_{\varepsilon, {\bf p},s}({\bf r},t)=\frac{e^{-i(
E^{(l)}_{\varepsilon}t-{\bf p}{\bf r})}}{2L^{\frac{3}{2}}}
\left(%
\begin{array}{c}{\sqrt{1+ \frac{m_l}{ E^{(l)}_
{\varepsilon}-c\alpha_n m_l}}} \ \sqrt{1+s\frac{p_{3}}{p}}
\\
{s \sqrt{1+ \frac{m_l}{ E^{(l)}_{\varepsilon}-c\alpha_n m_l}}} \
\sqrt{1-s\frac{p_{3}}{p}}\ \ e^{i\delta}
\\
{  s\varepsilon \eta\sqrt{1- \frac{m_l}{
E^{(l)}_{\varepsilon}-c\alpha_n m_l}}} \ \sqrt{1+s\frac{p_{3}}{p}}
\\
{\varepsilon \eta\sqrt{1- \frac{m_l}{ E^{(l)}_{\varepsilon}-c\alpha_n
m_l}}} \ \ \sqrt{1-s\frac{p_{3}}{p}}\ e^{i\delta}
\end{array}
\right).
\end{equation}
where the energy spectra are
\begin{equation}\label{Energy_e}
  E_{\varepsilon}^{(l)}=
  \varepsilon \eta \sqrt{{{\bf p}}^{2}\Big(1-s\alpha_n
  \frac{m_l}{p}\Big)^{2}
  +{m}^2} +c_l {\alpha}_n m_l, \qquad \alpha_n=\pm\frac{1}{2\sqrt{2}}
  {G_F}\frac{n_n}{m_l}.
\end{equation}
Here $p$, $s$ and $\varepsilon$ are the particles' momenta, helicities and signs of energy, ``$\pm$''
corresponds to $e$ and $\nu_e$. The value $\eta=$sign$\left(1-s\alpha_nm_l/p\right)$ is introduced to provide a
proper behavior of the neutrino wave function in a hypothetical massless case.

The developed approach to description of the matter effect on neutrinos and electrons, driven by (electro)weak
forces, is valid as long as interactions of particles with the background is coherent. This condition is
satisfied when a macroscopic amount of the background particles are confined within the scale of a neutrino or
electron de Broglie wave length. For relativistic neutrinos or electrons the following condition should be
satisfied $n/(\gamma_l m^3_l)\gg 1$, where $n$ is the number density of matter, $\gamma_l=E_l/m_l$ and $(l=\nu$
or $e )$. In a case of varying density of the background matter, there is an additional condition for
applicability of the developed approach (see, for instance, \cite{LoePRL90DvoStuJHEP02SilShuPP00}). The
characteristic length of matter density variations should be much larger than the de Broglie wavelength, $|\frac
{\bm \nabla n}{np}|\ll 1$.

Using the exact solutions of the above Dirac equations for a neutrino and an electron we have performed detailed
investigations of $SL\nu$ and $SLe$ in matter. In particular, in the case of ultra-relativistic neutrinos ($p\gg
m$) and a wide range of the matter density parameter $\alpha$ for the total rate of $SL\nu$ we obtained
\cite{StuTerPLB05GriStuTerPLB05GriStuTerG&C05}
\begin{equation}\label{gamma_nu}
\Gamma_{SL\nu} = 4 \mu_\nu ^2 \alpha ^2 m_\nu^2 p,  \ \ \ \ \ \ \ \
  {m_\nu}/{p} \ll \alpha \ll {p}/{m_\nu}.
\end{equation}
Performing the detailed study of the $SLe$ in neutron matter
\cite{GriShiStuTerTroG&C08_GriStuTerTroShiFizJ07} $^{,}$ \cite{}  we
have found for the total rate
\begin{equation}\label{gamma_e}
\Gamma_{SLe}=e^2 m_e^2/(2p)\left[\ln\big({4\alpha_n p}/{m_e}\big)-
    {3}/{2}\right],  \ \ \ \ {m_e}/{p}\ll\alpha_n\ll {p}/{m_e},
\end{equation}
where it is supposed that $\ln(4\alpha _n p/m_e) \gg 1$. It was also found that for relativistic electrons the
emitted photon energy can reach the range of gamma-rays. Furthermore, the electron can loose almost the whole
initial energy due to the $SLe$ mechanism.

Recently we apply our method to a particular case where a neutrino is propagating in a rotating medium of
constant density \cite{{GriSavStuRPJ07}} (see also \cite{StuJPA08,GriStuTerPAN09}). Suppose that a neutrino is
propagating perpendicular to the uniformly rotating matter composed of neutrons. This can be considered for
modelling of the neutrino propagation inside a rotating neutron star. The corresponding modified Dirac equation
for the neutrino wave function is given by (\ref{new_e}) with the matter potential accounting for rotation,
\begin{equation}\label{rot_f}
{\tilde f}^\mu = -{G}(n,n {\bf v}), \ \ {\bf v}=(\omega y,0,0),
\end{equation}
where $G=G_F/\sqrt{2}$. Here $\omega$ is the angular frequency of the matter rotation around the $z$ axis, here
also is accounted that all radial directions orthogonal to the $z$ axis are physically equal. For the energy of
the active left-handed neutrino we get
\begin{equation}\label{nu_quant_energy}
\label{energy_L} p_0 = \sqrt{p_3^2 + 2 \rho N} - G n, \qquad N=0,1,2,... \ .
\end{equation}
The energy depends on the neutrino momentum component $p_3$ along the rotation axis of matter and the quantum
number $N$ that determines the magnitude of the neutrino momentum in the orthogonal plane. For description of
antineutrinos one has to consider the ``negative sign'' energy eigeinvalues (for details see, for instance,
\cite{StuJPA08}). The energy of an electron antineutrino in the rotating matter composed of neutrons is given by
\begin{equation}\label{antinu_quant_energy}
\label{energy_L1} \tilde p_0 = \sqrt{p_3^2 + 2 \rho N} + G n, \qquad N=0,1,2,... \ .
\end{equation}
Thus, the transversal motion of the active neutrino and antineutrino is quantized in moving matter very much
alike an electron energy is quantized in a constant magnetic field that corresponds to the relativistic form of
the Landau energy levels (see, for instance,
 \cite{SokTerSynRad68}).

In conclusion of this section, we note that the developed new approach establishes a basis for investigation of
different phenomena which can emerge when neutrinos and electrons move in dense media, including those peculiar
to astrophysical and cosmological environments.

\section{Muon decay  $\mu\rightarrow e\nu\bar\nu$ in electromagnetic field}
In this section we inspect, using the method of exact solutions,
some aspects of the muon decay process in which the muon is embedded
in a field of a linearly polarized electromagnetic wave with the wave
vector $k=(\omega,{\bf k})$, where $\omega$ is the frequency and
$|{\bf k}|=\omega$. The field is thus described by the vector
potential $A(x)=a\cos(k\cdot x)$ satisfying the Lorenz condition,
where $a=(0,{\bf a})$ is a constant four-vector such that $a\cdot
k=0$.

The theoretical framework for the considered process was developed
in the basic papers of Ritus~\cite{ritus69,ritus87}. Its key
ingredients are the standard theory of the weak
interaction
and description of the muon and electron states by the Volkov functions~\cite{volkov35}. The decay rate is thus given
by~\cite{ritus69}
\begin{eqnarray}
W=\frac{G^2}{48\pi^4
q_0}\sum_{s>s_0}\int\frac{d^3q'}{q_0'}\left\{\left[\frac12(m_\mu^2-m_e^2)^2+\frac12(m_\mu^2+m_e^2)Q^2
-Q^4\right]B_0^2+\right. \nonumber\\
\left. +\left[\frac{(2Q^2+m_\mu^2+m_e^2)(k\cdot Q)^2}{(k\cdot
q)(k\cdot q')}+2Q^2\right]e^2a^2(B_1^2-B_0B_2)\right\},
\label{eq:rate}
\end{eqnarray}
where $G$ is the weak interaction constant,
$q=p-k(e^2a^2)/[4(k\cdot p)]$ and $q'=p'-k(e^2a^2)/[4(k\cdot p')]$
are the muon and electron four-quasimomenta ($p$ and $p'$ are the
field-free four-momenta, respectively), $m_\mu$ and $m_e$ are the
muon and electron masses, $s$ is the number of photons absorbed
from the wave (emitted into the wave if $s<0$),
$s_0=(m_\mu^2-m_e^2)/[2(k\cdot q)]$, $Q=sk+q-q'$. The functions
$B_i\equiv B_i(s;\alpha,\beta)$ ($i=0,1,2$), with
$$
\alpha=e\left(\frac{a\cdot p}{k\cdot p}-\frac{a\cdot p'}{k\cdot
p'}\right), \qquad \beta=\frac{e^2a^2}{8}\left(\frac{1}{k\cdot
p'}-\frac{1}{k\cdot p}\right),
$$
are defined as follows:
\begin{eqnarray}
B_0(s;\alpha,\beta)&=&\sum_{-\infty}^\infty
J_{s+2l}(\alpha)J_l(\beta),\nonumber\\
B_1(s;\alpha,\beta)&=&\frac12\left[B_0(s-1;\alpha,\beta)+B_0(s+1;\alpha,\beta)\right],\nonumber\\
B_2(s;\alpha,\beta)&=&\frac14\left[B_0(s-2;\alpha,\beta)+2B_0(s;\alpha,\beta)+B_0(s+2;\alpha,\beta)\right],
\end{eqnarray}
where $J_s$ is a Bessel function of order $s$. Note that practical calculations of Eq.~\ref{eq:rate} are
hindered by the sums and integrations involving rapidly oscillating functions.

We have reanalyzed the muon decay process in the electromagnetic field because recently Liu {\it et
al}~\cite{liu07} made a rather unexpected theoretical conclusion, based on their numerical calculations, that
the muon lifetime can be changed dramatically in an intense laser field achievable with present-day laser
sources. For example, they predicted an order-of-magnitude reduction of the muon lifetime in the case of a
CO$_2$ laser ($\omega=0.117$ eV) with the electric field amplitude $\mathcal{E}_0=\omega|{\bf a}|=10^6$ V/cm. It
should be remarked that Liu {\it et al}~\cite{liu07} employed instead of Eq.~\ref{eq:rate} an approximate model
which neglects the laser influence on the muon
and drops in the electron's Volkov state the dependence on terms
quadratic in the vector potential.
Such an approach as well as a resultant surprising prediction met
a serious criticism by Narozhny and Fedotov~\cite{narozhny08} who
classified the obtained result as ``fallacious'' and attributed it
either to the mistakes in analytics or to the erroneous numerical
calculation (see also the reply of Liu {\it et al}~\cite{liu08}).

Let us examine wether the result of Liu {\it et al}~\cite{liu07} can be inferred from Eq.~\ref{eq:rate} or not.
We note that, due to the properties of Bessel functions, the functions $B_0^2$ and $B_1^2-B_0B_2$ decrease
exponentially with $s$ if $|s|>\alpha$, and therefore we can replace the sum in Eq.~\ref{eq:rate} with
$\sum_{s=-\tilde{s}}^{\tilde{s}}$, where $\tilde{s}\sim ea/\omega$. For a CO$_2$ laser with $\mathcal{E}_0=10^6$
V/cm we have $\tilde{s}\sim2\cdot10^3$ ($|s_0|\sim5\cdot10^8$). Further, since all the items in the integrand of
Eq.~\ref{eq:rate} except $B_0^2$ and $B_1^2-B_0B_2$ practically do not vary with $s\leq\tilde{s}$, we can
following Ritus~\cite{ritus87} perform the summation over $s$ setting approximately $\tilde{s}=\infty$ and using
the formulas $\sum_{-\infty}^\infty B_0^2=1$ and $\sum_{-\infty}^\infty(B_1^2-B_0B_2)=0$. The remaining
integrations can be performed analytically and they yield the well known vacuum result, i.e. $W\approx
W_0=G^2m_\mu^5/(192\pi^3)$. We can also estimate the relative change in the decay rate as
$$
\delta=\frac{W-W_0}{W_0}\simeq\frac{37}{48}\frac{e^2a^2}{m_\mu^2}\sim-10^{-12}
$$
which is due to the difference between the muon and electron
four-quasimomenta, $q$ and $q'$, and their field-free analogs, $p$
and $p'$. This estimate clearly shows the nonphysical character of
the prediction made by Liu {\it et al}~\cite{liu07}.


\section*{Acknowledgments}
 One of us (A.S.) would like to thank Jean Tran Thanh Van and Jacques
Dumarchez for the invitation to attend the XXth Rencontres de Blois on Challenges in Particle Astrophysics  and
all of the organizers of this conference for their warm hospitality. We are grateful to Massimo Passera who draw
our attention to the Letter \cite{liu07} devoted to the problem of the muon decay in a laser field and to
Vladimir Ritus for very useful discussions on this issue.

\section*{References}

\end{document}